%% file: main.tex
\documentclass[conference]{IEEEtran}
\IEEEoverridecommandlockouts

\usepackage{orcidlink}
\usepackage{cite}
\usepackage{amsmath,amssymb,amsfonts}
\usepackage{algorithmic}
\usepackage{graphicx}
\usepackage{textcomp}
\usepackage{xcolor}
\usepackage{dsfont}

\usepackage[nolist]{acronym}
\usepackage{xcolor}
\usepackage{comment}
\usepackage{cite}
\usepackage{soul}

\usepackage{subcaption}
\usepackage{epstopdf}
\usepackage{xfrac}
\usepackage{pifont}

\usepackage{csquotes}
\usepackage{booktabs}
\usepackage{multirow}
\usepackage{tabularx,ragged2e}
\newcolumntype{Y}{>{\centering\arraybackslash}X}
\usepackage{footmisc}
\usepackage{hyperref}
\usepackage{cleveref}

\usepackage{svg}
\usepackage{varwidth}
\usepackage{tikz}
\usepackage{pgf}
\usepackage{pgfplots}
\usepackage{pgfplotstable}
\usetikzlibrary{positioning,backgrounds,fit,calc,patterns,arrows.meta}
\pgfplotsset{compat=1.3}
\usetikzlibrary{shapes.geometric}
\usetikzlibrary{shapes.arrows}
\usetikzlibrary{arrows.meta}
\usepackage[skins]{tcolorbox}
\usepackage{adjustbox}

\usepackage{array}

\newcommand{\hrtf}[0]{\mathbf{a}} %
\newcommand{\speech}[0]{\mathbf{s}}
\newcommand{\rir}[0]{\mathbf{h}}
\newcommand{\reverb}[0]{\mathbf{y}}
\newcommand{\doa}[0]{\gamma}

\newcommand{\D}[0]{\mathrm{d}}
\newcommand{\state}[0]{\hrtf_\tau}

\newcommand{\clean}[0]{\hrtf_0}
\newcommand{\init}[0]{\hrtf_{T_\mathrm{max}}}
\newcommand{\sco}[0]{\nabla_{\state} \log p(\state|\gamma)}

\newcommand{\postsco}[0]{\nabla_{\state} \log p(\state | \reverb, \speech, \gamma)}
\newcommand{\likelihood}[0]{\nabla_{\state} \log p(\reverb | \state, \speech)}
\newcommand{\scomodel}[0]{\mathbf{s}_\theta(\state, \tau, \gamma)}

\makeatletter
\newcommand{\subalign}[1]{%
  \vcenter{%
    \Let@ \restore@math@cr \default@tag
    \baselineskip\fontdimen10 \scriptfont\tw@
    \advance\baselineskip\fontdimen12 \scriptfont\tw@
    \lineskip\thr@@\fontdimen8 \scriptfont\thr@@
    \lineskiplimit\lineskip
    \ialign{\hfil$\m@th\scriptstyle##$&$\m@th\scriptstyle{}##$\hfil\crcr
      #1\crcr
    }%
  }%
}
\makeatother

\begin{document}

\title{
HRTF Estimation using a Score-based Prior \\
}

\author{\IEEEauthorblockN{Etienne Thuillier}
\IEEEauthorblockA{\textit{Acoustics Lab} \\
\textit{Aalto University}\\
Espoo, Finland \\
\orcidlink{0009-0007-1619-1947}}
\and
\IEEEauthorblockN{Jean-Marie Lemercier}
\IEEEauthorblockA{\textit{Signal Processing Group} \\
\textit{University of Hamburg}\\
Hamburg, Germany \\
\orcidlink{0000-0002-8704-7658}}
\and
\IEEEauthorblockN{Eloi Moliner}
\IEEEauthorblockA{\textit{Acoustics Lab} \\
\textit{Aalto University}\\
Espoo, Finland \\
\orcidlink{0000-0001-5719-326X}}
\and
\IEEEauthorblockN{Timo Gerkmann}
\IEEEauthorblockA{\textit{Signal Processing Group} \\
\textit{University of Hamburg}\\
Hamburg, Germany \\
\orcidlink{0000-0002-8678-4699}}
\and
\IEEEauthorblockN{Vesa V\"alim\"aki}
\IEEEauthorblockA{\textit{Acoustics Lab} \\
\textit{Aalto University}\\
Espoo, Finland \\
\orcidlink{0000-0002-7869-292X}}
}

\maketitle

\begin{abstract}
We present a head-related transfer function (HRTF) estimation method which relies on a data-driven prior given by a score-based diffusion model.
The HRTF is estimated in reverberant environments using natural excitation signals, e.g. human speech. 
The impulse response of the room is estimated along with the HRTF by optimizing a parametric model of reverberation based on the statistical behaviour of room acoustics.
The posterior distribution of HRTF given the reverberant measurement and excitation signal is modelled using the score-based HRTF prior and a log-likelihood approximation.
We show that the resulting method outperforms several baselines, including an oracle recommender system that assigns the optimal HRTF in our training set based on the smallest distance to the true HRTF at the given direction of arrival.
In particular, we show that the diffusion prior can account for the large variability of high-frequency content in HRTFs.

\end{abstract}


\begin{acronym}
\acro{stft}[STFT]{short-time Fourier transform}
\acro{istft}[iSTFT]{inverse short-time Fourier transform}
\acro{dnn}[DNN]{deep neural network}
\acro{pesq}[PESQ]{Perceptual Evaluation of Speech Quality}
\acro{polqa}[POLQA]{perceptual objectve listening quality analysis}
\acro{wpe}[WPE]{weighted prediction error}
\acro{psd}[PSD]{power spectral density}
\acro{rir}[RIR]{room impulse response}
\acro{hrir}[HRIR]{head-related impulse response}
\acro{brir}[BRIR]{binaural room impulse response}
\acro{hrtf}[HRTF]{head-related transfer function}
\acro{snr}[SNR]{signal-to-noise ratio}
\acro{lstm}[LSTM]{long short-term memory}
\acro{polqa}[POLQA]{Perceptual Objective Listening Quality Analysis}
\acro{sdr}[SDR]{signal-to-distortion ratio}
\acro{estoi}[ESTOI]{extended short-term objective intelligibility}
\acro{elr}[ELR]{early-to-late reverberation ratio}
\acro{tcn}[TCN]{temporal convolutional network}
\acro{rls}[RLS]{recursive least squares}
\acro{asr}[ASR]{automatic speech recognition}
\acro{ha}[HA]{hearing aid}
\acro{ci}[CI]{cochlear implant}
\acro{mac}[MAC]{multiply-and-accumulate}
\acro{vae}[VAE]{variational auto-encoder}
\acro{gan}[GAN]{generative adversarial network}
\acro{tf}[T-F]{time-frequency}
\acro{sde}[SDE]{stochastic differential equation}
\acro{ode}[ODE]{ordinary differential equation}
\acro{drr}[DRR]{direct to reverberant ratio}
\acro{lsd}[LSD]{log spectral distance}
\acro{sisdr}[SI-SDR]{scale-invariant signal to distortion ratio}
\acro{mos}[MOS]{mean opinion score}
\acro{map}[MAP]{maximum a posteriori}
\acro{sde}[SDE]{stochastic differential equation}
\acro{ode}[ODE]{ordinary differential equation}
\acro{dps}[DPS]{diffusion posterior sampling}
\acro{fir}[FIR]{finite impulse response}
\acro{doa}[DoA]{direction of arrival}
\acro{ild}[ILD]{inter-aural level difference}
\acro{itd}[ITD]{inter-aural time difference}
\acro{ipd}[IPD]{inter-aural phase difference}
\end{acronym}

\section{Introduction}
\label{sec:intro}
\input{sections/introduction}

\section{Score-based Diffusion Model of HRTFs}
\input{sections/sgm}

\section{Binaural Room Impulse Response Parameterization}
\input{sections/method}

\begin{figure}
    \centering
    \input{figures/diagram_small}
    \caption{Diagram of the inference algorithm.}
    \label{fig:diagram}
\end{figure}
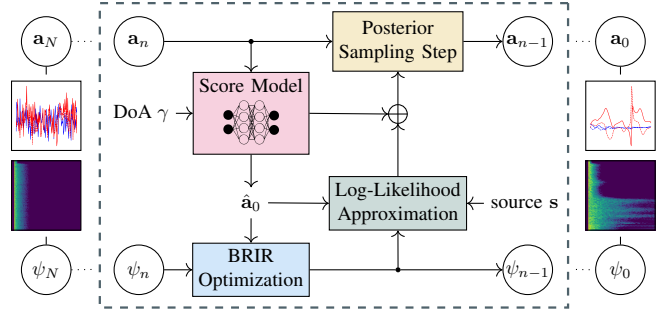

\section{Inference Algorithm}

\input{sections/inference}

\section{Experimental Setup}
\label{sec:exp}
\input{sections/exp}

\input{figures/plots_v3}

\section{Results and Discussion}
\label{sec:results}

\input{sections/results}

\section{Conclusion}

This paper proposed a posterior sampling scheme for HRTF estimation using a diffusion-based prior and a parameteric reverberation model for approximating log-likelihood computation.
Compared to previous approaches, the proposed method can use natural-sounding sources such as speech, requires only one measurement, and more importantly, it can operate in a variety of reverberant environments.
At the exception of target directions in the median plane, our method largely outperforms the considered baselines, including an oracle recommender system. 
In relative terms, the time-aligned HRTF estimation error is particularly low in the high-frequency region, which we attribute to the expressivity of the diffusion prior.

\bibliographystyle{IEEEtran}
\bibliography{bibliography}

\end{document}

%% file: sections/introduction.tex
Standard \ac{hrtf} measurements follow a system identification approach whereby a synthetic probe signal is rendered through a loudspeaker, picked-up by microphones located at the entrances of the subject’s ear canals, and used to deconvolve the resulting binaural recording~\cite{algazi2001cipic}.
Repeating the measurement to capture a full \ac{hrtf} is laborious and time-consuming.
The procedure also requires a dedicated anechoic chamber and specialized, calibrated, audio equipment.
Methods using non-specialized equipment in echoic environments have been proposed to democratize access to individualized~\acp{hrtf},
for example, using a living room's loudspeaker emitting short bursts of exponential sine sweeps~\cite{reijniers2020hrtf} or a hand-held smartphone emitting synthetic probe signals in near field~\cite{yang2021personalizing}.
While scalable to the mass-market and cost-effective, these approaches require the subjects to actively undergo a procedure during which they are subjected to unpleasant-sounding synthetic signals.

This issue could be addressed by relying on isolated sound sources occurring in the subject's surrounding environment.
Recently, Jayaram et al.~\cite{jayaram2023hrtf} trained a neural network in a supervised fashion to predict the \ac{hrtf}'s magnitude spectrum using recordings captured from consumer-grade binaural microphones.
This method relies on detection of the sound source's location and composes a full \ac{hrtf} by aggregating estimates from various \acp{doa}.
Such an approach could prove particularly advantageous if shown adaptable to modern earbud headphones, in which microphones are typically offset from the entrances of the ear canals.

In this work, we also propose to leverage binaural recordings of a source in the subject's everyday environment.
However, we suggest using sounds played back by the user over a paired device with known directivity patterns, for example podcast content over a smart speaker, such that the source is known a-priori along with its \ac{doa}.
This setup leads us to formulating the task of HRTF estimation from a blind inverse problem perspective, i.e. sampling a valid HRTF that is consistent with the observed binaural reverberant measurement, while also estimating the reverberation in the room.
The HRTF sampling procedure uses a data-driven prior provided by a diffusion model trained on binaural time-aligned \ac{hrtf} filter data.
We then fit the room acoustics to a parametric model adapted from \cite{moliner2024buddy, lemercier2024buddyjournal}, which we jointly optimize during the HRTF estimation.
This follows a recent line of work that applies diffusion models as priors to solve inverse problems in image and audio domains \cite{chung_diffusion_2022, moliner_solving_2022, lemercier2024buddyjournal}.

\par Unlike~\cite{jayaram2023hrtf}, the proposed approach recovers both magnitude and phase estimates. %
In contrast to a previous signal processing method under a similar
setup~\cite{durkovicd2010hrtf}, the prior ensures consistent scaling between measurements. 
Furthermore, we demonstrate that our approach outperforms a nearest-neighbour oracle baseline which returns the \ac{hrtf} from the training set that is closest to the true \ac{hrtf} at the the detected \ac{doa}.
In particular, the diffusion prior shows a good expressivity in the higher-frequency regions where the \ac{hrtf} presents large variations across subjects.
The size of the diffusion model is modest, potentially allowing for on-device processing.%

%% file: sections/sgm.tex
We present here our data-driven prior for time-aligned HRTF features 
based on continuous-time diffusion-based generative models, also known as score-based models.
Score-based models \cite{ho2020denoising, song2019generative} encompass a class of generative models particularly successful at learning complex data distributions such as e.g. natural images or human speech. 
Here, we employ score-based models to approximate $p(\mathbf{a} | \gamma)$, that is 
the distribution of time-aligned HRTFs in the frequency domain, denoted as 
$\hrtf \in \mathds{C}^{2 \times F}$
, conditioned on the \ac{doa} $\gamma$.
 
Score-based models operate as iterative Gaussian denoisers: during training, the target data distribution is transformed into a standard Gaussian distribution following a \textit{forward diffusion process}, incrementally adding noise. 
Once training is achieved, new data belonging to the data distribution can be generated through the \textit{reverse diffusion process}, which iteratively removes noise from an initial Gaussian sample until a data sample emerges.
In continuous-time score-based models 
\cite{song2019generative}, 
this reverse process
can be characterized by the following \emph{probability flow} \ac{ode} \cite{Oksendal2000SDE}, adopting the parameterization by Karras et al. \cite{karras2022elucidating}:
\begin{equation}\label{eq:ode}
    \D \state = - \tau \sco \D \tau, 
\end{equation}
where $\tau$ indexes the reverse process 
flowing from $T_\mathrm{max}$ to $0$. 
The diffusion state $\state$ starts from the initial condition $\init \sim \mathcal{N}(0,\sigma(T_\mathrm{max})^2 \mathbf{I})$ and terminates at $\clean \sim p_\text{data}$. 
We choose a linear noise variance schedule $\sigma(\tau)=\tau$,
which defines the  Gaussian marginal densities
$p_\tau(\state | \clean) = \mathcal{N}(\state; \clean, \sigma(\tau)^2 \mathbf{I})$.
The \emph{score} 
$\sco$ is intractable at inference for complex distributions.
Therefore, a \textit{score model} parameterized with a \ac{dnn} $\scomodel$ is trained to estimate the score
using \emph{denoising score matching} \cite{vincent2011connection}.

%% file: sections/method.tex
\par In a static setup where source and receiver locations are fixed, binaural reverberation 
can be modelled
by convolving an anechoic source $\speech$ 
with the impulse response of the system, i.e. the \ac{brir} $\rir$.
The BRIR is composed from the contributions of 
wavefronts traveling from the source to the ears of the subject following direct and indirect propagation paths.
In this work, we model the \ac{brir} as the sum of an anechoic and a reverberant component:
\begin{equation}
    \rir^{(\psi)}_\hrtf := \left[\begin{array}{c}\boldsymbol{\delta}_{t_{\text{left}}}\\\boldsymbol{\delta}_{t_{\text{left}}}\end{array}\right] \ast\left(g\left[\begin{array}{c} \boldsymbol{\delta}_0 \\ \boldsymbol{\delta}_{t_{\text{itd}}} \end{array}\right] \ast \mathcal{F}^{-1}(\hrtf) + \mathbf{r}^{(\chi)}\right),
    \label{eq:brir_hat}
\end{equation}
where
$\psi=\left\{t_{\text{left}},g,t_{\text{itd}}\right\} \cup \chi$ denotes the optimizable parameters of the model, and 
$\hrtf$ 
is the binaural time-aligned \ac{hrtf}~\cite{ben2019efficient} at the given \ac{doa}. The HRTF is defined in the frequency domain, hence its time-domain equivalent called head-related impulse response is obtained via inverse Fourier transform $\mathcal{F}^{-1}$.
The symbol $\boldsymbol{\delta}_t$ denotes a Kronecker delta delayed by $t$ seconds,
$t_{\text{left}} \in [0,\infty)$
models time of flight to the subject's left ear,
$g\in (0, 1]$ represents a real-valued attenuation constant,
$t_{\text{itd}}\in \mathds{R}$ denotes the \ac{itd} 
taking the left ear as the reference (non-delayed) channel,
and $\mathbf{r} \in \mathds{R}^{2\times N_r} $ denotes the reverberant component of the model.

The reverberant component $\mathbf{r}$ is 
implemented as a binaural variant of a previously proposed parametric model fitting the reverberant characteristics of the room~\cite{moliner2024buddy, lemercier2024buddyjournal}:
\begin{equation} \label{eq:brir-model}
\begin{array}{@{}r@{}c@{}l@{}}
    \mathbf{r}^{(\chi)} &:=& \mathcal{F}_{ST}^{-1}\left( \mathrm{interp.} \left( \left[w_b \operatorname{e}^{- m \alpha_b} \right]_{(c, m,b)} \right) \odot \left[\operatorname{e}^{j \phi_{c, m, f}} \right]_{(c, m,f)} \right),
\end{array}
\end{equation}
where
$\chi=\left\{\left[\alpha_{b}\right]_{b},\left[w_{b}\right]_{b}, \left[\phi_{c, m, f}\right]_{c, m, f}\right\}$ denotes the learnable parameters of $\mathbf{r}$,
$\mathcal{F}^{-1}_{ST}$ denotes the inverse \ac{stft}
, $\odot$ denotes the Hadamard element-wise product,
$m\in \left\{1,\hdots, M_\mathbf{r}\right\}$ indexes over \ac{stft} frames,
$f\in \left\{1,\hdots, F\right\}$ over Fourier bins,
$b\in \left\{1,\hdots, B\right\}$ over frequency subbands (with $B < F)$,
and $c\in\left\{ \text{left}, \text{right} \right\}$.

The weight $w_b \in [0,\infty)$ and decay rate $\alpha_b \in [0, \infty)$ define a subband exponential decaying magnitude model, which exploits the observed statistical nature of reverberation tails \cite{Habets2010}.
Crucially, this magnitude envelope is
identical for the left and right channels.
The interpolation operator $\mathrm{interp.}$ upsamples the $B$ subbands to the $F$ frequency bins of the STFT by employing $\exp(\mathrm{lerp}(\log(\cdot)))$, where $\mathrm{lerp}$ denotes linear interpolation.
In contrast to the magnitude envelope term, the phase is determined using a specific coefficient $\phi_{c,m,f}\in\left[\pi,\pi\right)$ for each channel-frame-bin triplet.
This allows for fitting decorrelated left and right channel realizations of the diffuse part, as typically observed in  \acp{brir} above 1 kHz \cite{menzer2009investigations,fagerstrom2024binaural}.
Additionally, while the early reflections of the BRIR are not explicitly incorporated into the model, the unconstrained phases allow the model to fit these reflections to some extent.

%% file: figures/diagram_small.tex
\definecolor{cb1}{HTML}{D81B60}
\definecolor{cb2}{HTML}{1E88E5}
\definecolor{cb3}{HTML}{D29E02}
\definecolor{cb4}{HTML}{004D40}
\definecolor{cb5}{HTML}{864dbf}

\tikzstyle{mycircle} = [circle, draw, fill=white, inner sep=0pt, minimum size=25pt]
\tikzstyle{mysquare} = [rectangle, draw, fill=black, inner sep=0.1pt, minimum width=35pt, minimum height=35pt, align=center]
\tikzstyle{mybranch} = [circle, draw, fill=black, inner sep=0pt, , minimum size=2pt]
\tikzstyle{myrectangle} = [rectangle, draw, fill=cb1!20, inner sep=3pt, minimum width=35pt, minimum height=20pt, align=center]
\tikzstyle{myrectangle2} = [rectangle, draw, fill=cb2!20, inner sep=3pt, minimum width=50pt, minimum height=20pt, align=center]
\tikzstyle{myrectangle4} = [rectangle, draw, fill=cb3!20, inner sep=3pt, minimum width=50pt, minimum height=30pt, align=center]
\tikzstyle{myrectangle3} = [rectangle, draw, fill=cb4!20, inner sep=3pt, minimum width=30pt, minimum height=25pt, align=center]
\tikzstyle{myrectangle5} = [rectangle, draw, fill=cb5!20, inner sep=3pt, minimum width=30pt, minimum height=25pt, align=center]

\tikzstyle{myrectangle_white} = [rectangle, draw, fill=white, inner sep=3pt, minimum width=10em, minimum height=20pt, align=center]

\tikzstyle{sum} = [
  circle,
  draw,
  minimum size=9pt,
  append after command={
    \pgfextra{
      \draw (\tikzlastnode.north) -- (\tikzlastnode.south);
      \draw (\tikzlastnode.west) -- (\tikzlastnode.east);
    }
  },
]

\tikzstyle{product} = [
  circle,
  draw,
  minimum size=12pt,
  append after command={
    \pgfextra{
      \draw (\tikzlastnode.north east) -- (\tikzlastnode.south west);
      \draw (\tikzlastnode.north west) -- (\tikzlastnode.south east);
    }
  },
]

\tikzstyle{convolutioncircle} = [
  circle,
  draw,
  minimum size=24pt,
]

\newcommand{\specwidth}{1.35cm}
\newcommand{\sepnodex}{0.5cm}
\newcommand{\sepnodey}{0.4cm}
\newcommand{\sepnodexspec}{0cm}
\newcommand{\sepnodeyspec}{0.25cm}
\newcommand{\sepstate}{0.75cm}
\definecolor{figblue}{HTML}{154c79}
\definecolor{figgreen}{HTML}{a8db33}
\definecolor{figgrey}{HTML}{797979}

\definecolor{uhhstone}{RGB}{59,81,91}

\begin{tikzpicture}[scale=0.75, transform shape]

\coordinate (stateN) at (0,0.0);
\coordinate (statepsiN) at (0,-4.05);

\node[rectangle, draw, color=uhhstone!90!white, dashed, line width=0.3mm, minimum width=8.3cm, minimum height=5.4cm] at (5.1, -2.02) (rect) {};

\node[mycircle] (sN) at (stateN) {$\mathbf{a}_{N}$};
\node[mycircle] (spsiN) at (statepsiN) {$\psi_{N}$};
\node[mysquare, below=\sepnodeyspec of sN, anchor=north, fill overzoom image={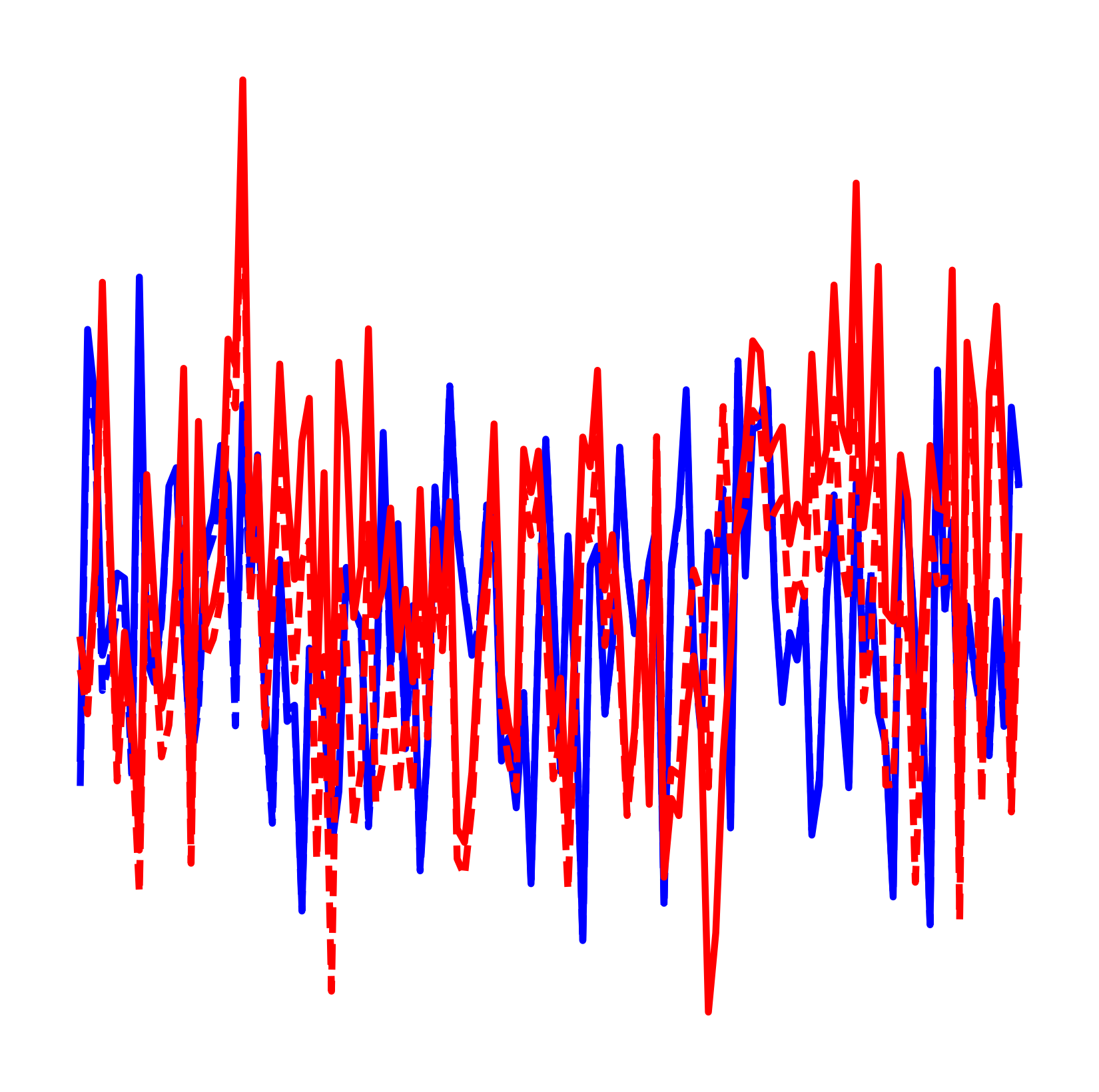}] (sNspec) {};
\draw[-, dashed] (sN) to (sNspec);
\node[mysquare, above=\sepnodeyspec of spsiN, anchor=south, fill overzoom image={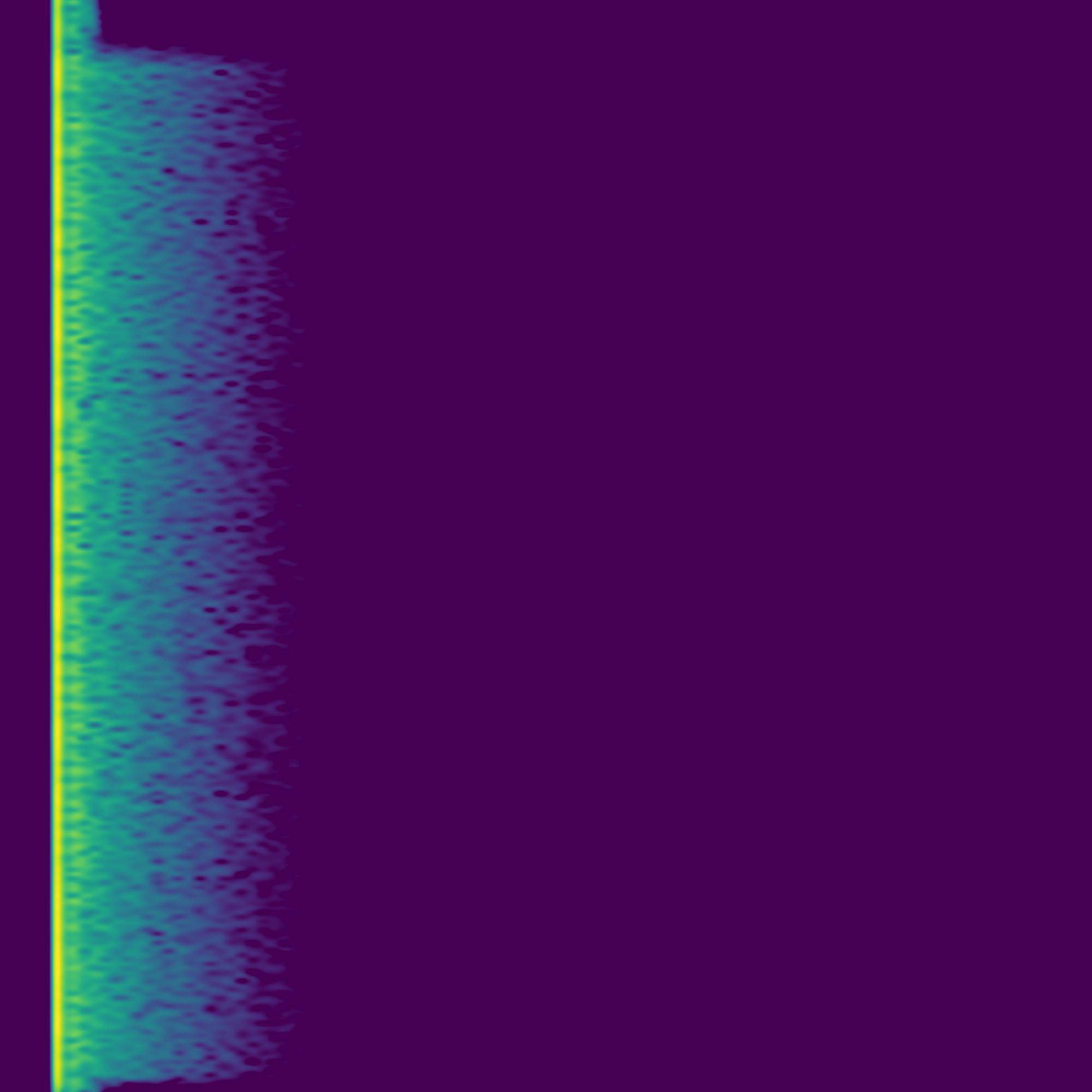}] (spsiNspec) {};
\draw[-, dashed] (spsiN) to (spsiNspec);

\node[mycircle, right=0.75cm of sN] (sn) {$\mathbf{a}_{n}$};
\node[mycircle, right=0.75cm of spsiN] (spsin) {$\psi_{n}$};
\node[above=1.898cm of rect.west] (block_input_sn) {};
\draw[-, dotted] (sN.east) to (block_input_sn);
\node[below=1.898cm of rect.west] (block_input_spsin) {};
\draw[-, dotted] (spsiN.east) to (block_input_spsin);

\node[mybranch, right=1.5 of sn] (branchsn) {};
\node[myrectangle, align=center, below=0.5cm of branchsn] (score) {Score Model
 \\ \vspace{-0.3cm}\\ \includegraphics[height=25pt]{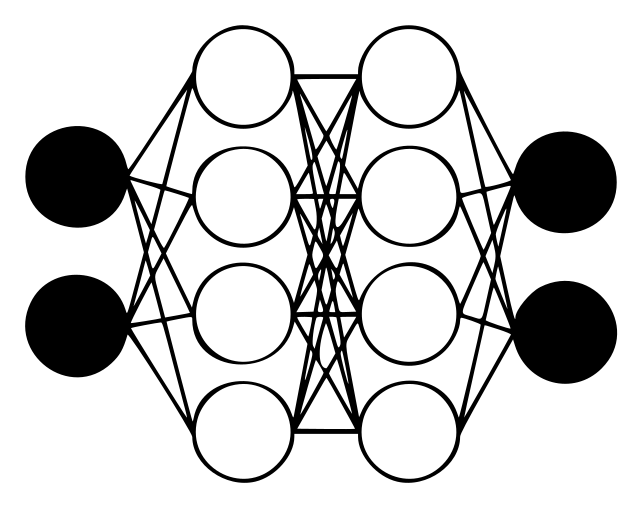}
 };
\node[below=0.58cm of score] (x0) {$\hat{\mathbf{a}}_0$};
\draw[->] (branchsn) -- (score.north);
\draw[->] (score.south) -- (x0.north);

\node[below=0.2cm of score.west] (input_doa) {};
\node[left=0.3cm of score.west] (signal_doa) {DoA\;$\gamma$};
\draw[->] (signal_doa) -- (score.west);

\node[myrectangle2, align=center, right=0.5cm of spsin] (rir) {BRIR\\Optimization};

\node[mybranch, right=1.525cm of rir] (branch_rir) {};
\draw[->] (spsin) -- (rir);
\draw[->] (x0) -- (rir);
\draw[-] (rir) -- (branch_rir);
\node[myrectangle4, align=center, right=1.4cm of branchsn] (dps) {Posterior\\Sampling Step};
\node[sum, below=0.58cm of dps] (sum) {};
\draw[->] (sn.east) -- (dps.west);
\node[myrectangle3, below=0.93 of sum] (distance) {Log-Likelihood\\Approximation};

\node[right=0.3cm of distance.east] (signal_s) {source $\speech$};
\draw[->] (signal_s) -- (distance.east);

\draw[->] (distance.north) -- (sum.south);
\draw[->] (sum.north) -- (dps.south);
\draw[->] (score.east) -- (sum.west);
\draw[->] (branch_rir) -- (distance.south);
\draw[->] (x0.east) -- (distance.west);

\node[mycircle, right=6cm of sn] (snminus1)  {$\mathbf{a}_{n-1}$};
\draw[->] (dps.east) -- (snminus1.west);
\node[mycircle, right=6cm of spsin] (spsinminus1)  {$\psi_{n-1}$};
\draw[->] (branch_rir) -- (spsinminus1.west);

\node[mycircle, right=0.75cm of snminus1] (s0) {$\mathbf{a}_{0}$};
\node[mycircle, right=0.75cm of spsinminus1] (spsi0) {$\psi_{0}$};
\node[mysquare,  below=\sepnodeyspec of s0, anchor=north, fill overzoom image={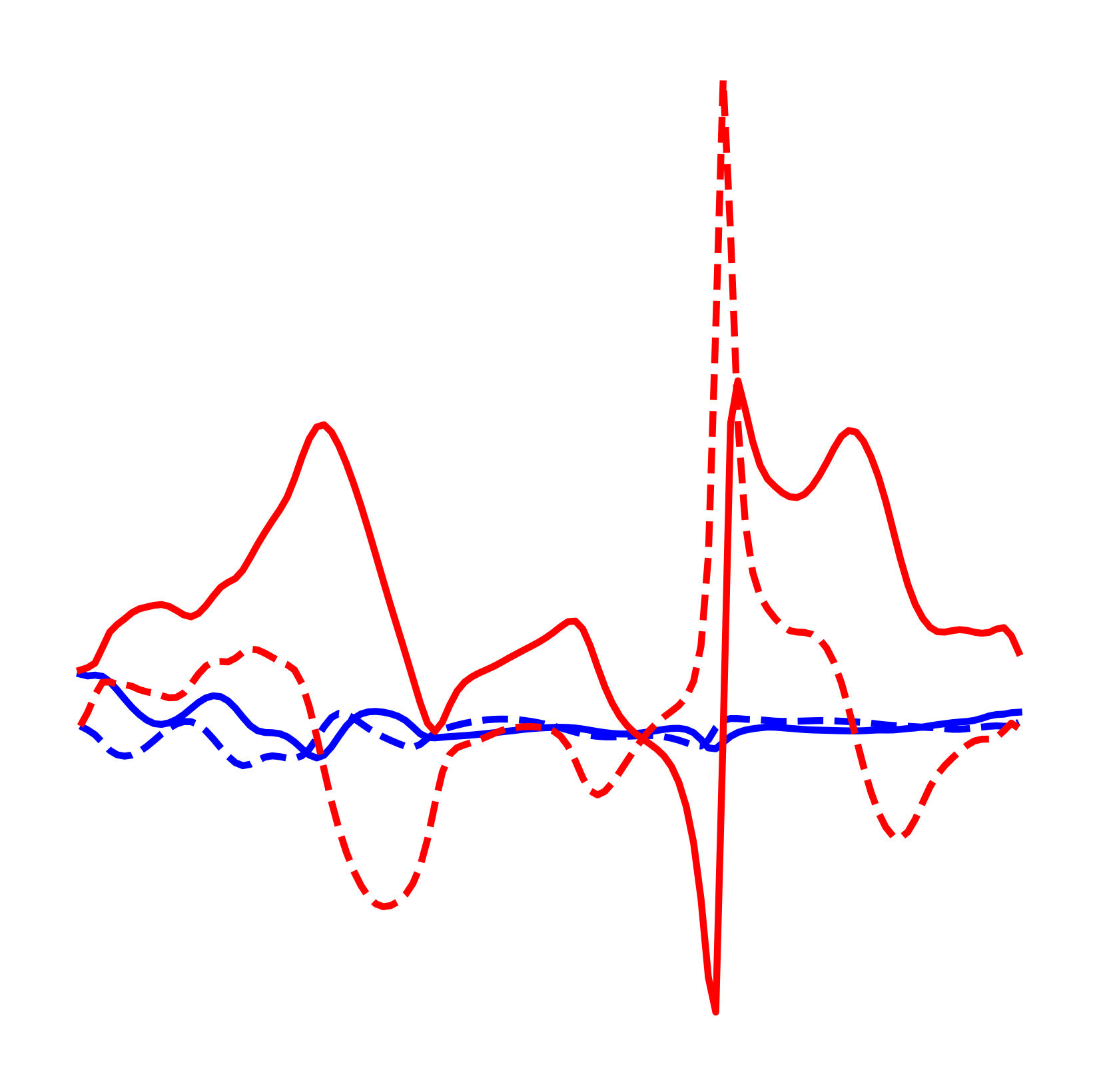}] (s0spec) {};
\draw[-, dashed] (s0) to (s0spec);
\node[mysquare,  above=\sepnodeyspec of spsi0, anchor=south, fill overzoom image={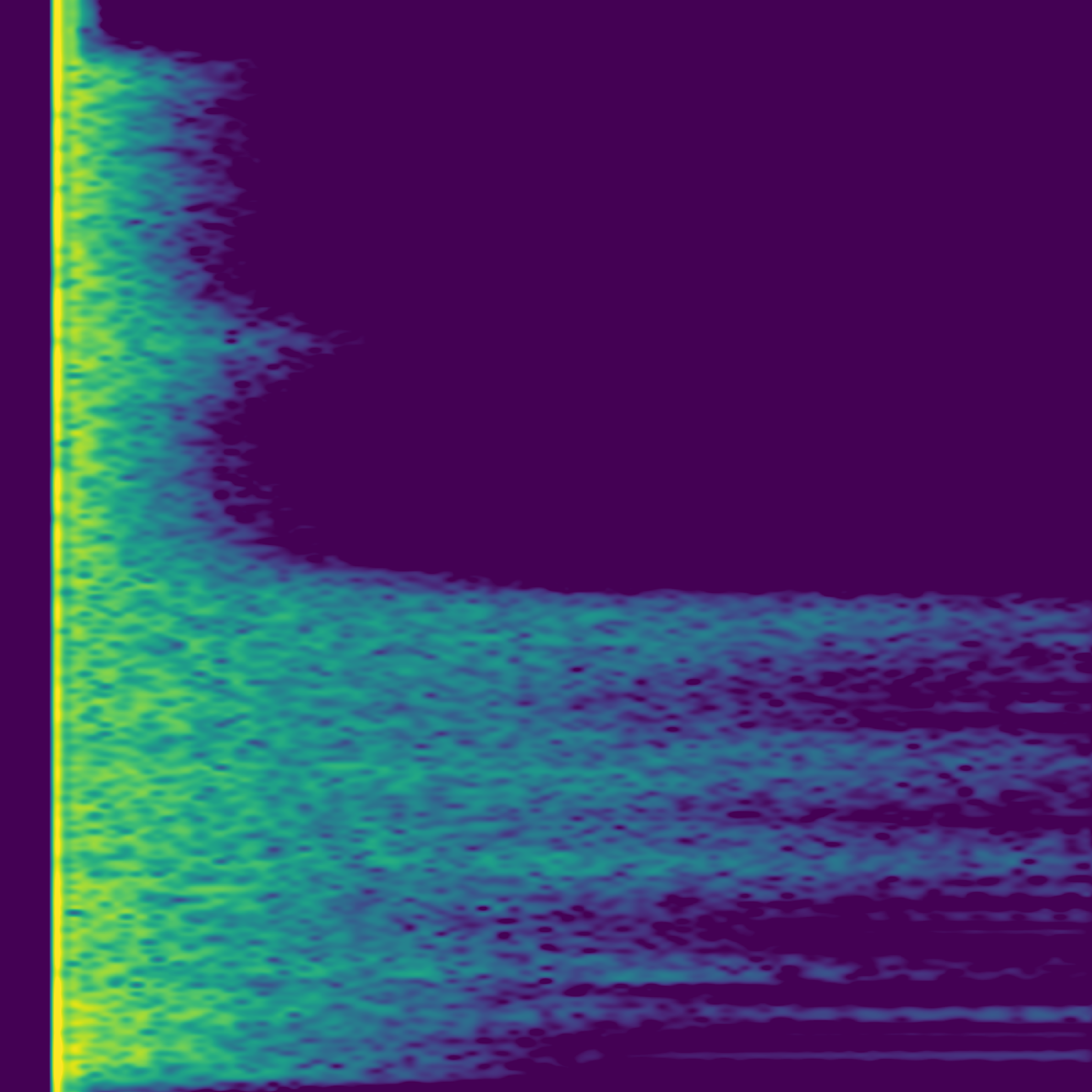}] (spsi0spec) {};
\draw[-, dashed] (spsi0) to (spsi0spec);

\node[above=1.898cm of rect.east] (block_output_sn) {};
\draw[-, dotted] (block_output_sn) to (s0.west);
\node[below=1.898cm of rect.east] (block_output_spsin) {};
\draw[-, dotted] (block_output_spsin) to (spsi0.west);

\end{tikzpicture}

%% file: sections/inference.tex
The inference process solves the following objective:
\begin{equation}
\label{eq:inverse_problem_optim}
  \hat{\hrtf}, \hat{\psi} = \underset{\hrtf, \psi}{\mathrm{arg\,min}}\;  
    \mathcal{C}(\mathbf{y},  \rir^{(\psi)}_\hrtf \ast \speech  ) \;\;\;
     \text{s.t.}\;\;\; \hrtf \sim p_{\text{data}}
\end{equation}

\noindent This means that we wish to retrieve the optimal time-aligned \ac{hrtf} $\hat{\hrtf}$ and \ac{brir} parameters $\hat{\psi}$ in order to minimize a reconstruction error $\mathcal{C}(\mathbf{y},  \rir^{(\psi)}_\hrtf \ast \speech  ) $ given and the binaural reverberant measurement $\mathbf{y}$ and the broadband excitation signal $\speech$, which is in our case (but not restricted to) human speech.
Another soft constraint is that the estimated time-aligned \ac{hrtf} $\hat{\hrtf}$ should belong to the target \ac{hrtf} distribution $p_{\text{data}}$.

We solve \eqref{eq:inverse_problem_optim} using an alternating optimization procedure, visualized in Fig.~\ref{fig:diagram}.
The acoustic parameters $\hat{\psi}$ in our \ac{brir} model \eqref{eq:brir-model} are updated with a classical gradient-based optimizer (e.g. Adam) minimizing the reconstruction loss.
The \ac{hrtf} estimate $\hat{\hrtf}$, however, is refined using a posterior sampling technique leveraging a score model $\mathbf{s}_\theta(\state, \tau, \doa)$ trained on time-aligned \ac{hrtf} data and conditioned on the \ac{doa} $\doa$.
Precisely, this procedure aims at sampling from the posterior distribution 
$p(\hrtf | \reverb, \speech, \gamma)$.
We leverage our \ac{hrtf} score-based prior for posterior sampling by solving the ODE \eqref{eq:ode}, where  the score function $\sco$ is replaced by the so-called \textit{posterior score} obtained via Bayes' formula:
\begin{equation} \label{eq:bayes}
    \postsco = \sco + \likelihood \, .
\end{equation}
The \emph{prior score} $\sco$ is obtained via our score model
$\scomodel$.
Since we generally do not have a model for $\reverb$ given the diffusion state $\state$, the likelihood 
$p(\reverb | \state, \speech)$ 
is intractable.
However, it can be derived using the following approximations, similar to \cite{moliner2024buddy}.
First, we follow Chung et al. \cite{chung_diffusion_2022} and employ an estimate of $\hrtf_0$ at time $\tau$ as a sufficient statistic for $\state$ in the likelihood. This results in assuming $p(\reverb | \state, \speech) \approx p(\reverb | \hat{\hrtf}_0, \speech)$. 
The estimate $\hat{\mathbf{a}}_0$ is directly obtained from the unconditional score model via one-step denoising:
\begin{equation}
    \hat{\mathbf{a}}_0 := 
    \state - \sigma(\tau)^2 \scomodel.
\end{equation}

Furthermore, we model binaural reverberation by a convolution between the excitation signal $\speech$ and our \ac{brir} model $\rir_{\hat{\hrtf}_0}^{(\psi)}$, using the one-step denoising \ac{hrtf} estimate $\hat{\hrtf}_0$,
and
assuming $p(\reverb | \hat{\hrtf}_0, \speech) \approx p(\reverb | \rir^{(\psi)}_{\hat{\hrtf}_0} \ast \speech ) $. 
Finally, we follow \cite{moliner2024blind} and approximate the log-likelihood gradient using a $L^2$-distance in the magnitude-compressed STFT domain, between the measurement and our estimate:
\begin{equation}\label{eq:objective}
\mathcal{C}(\mathbf{y}, \hat{\mathbf{y}})
= \frac{1}{M_\mathbf{y} }
\rVert 
S_\text{comp}(\mathbf{y})-S_\text{comp}(\hat{\mathbf{y}})
\lVert_2^2 \,,
\end{equation}
where $S_\text{comp}(\mathbf{y}) := |\mathcal{F}_{\text{ST}}(\mathbf{y})|^{2/3} \exp{j \angle \mathcal{F}_{\text{ST}}(\mathbf{y})}$ is the magnitude-compressed spectrogram with $M_\mathbf{y}$ \ac{stft} frames.
This compression accounts for the heavy-tailedness of speech distributions \cite{gerkmann2010}.
Note that we also use this function as the objective for optimizing the \ac{brir} parameters $\psi$ in \eqref{eq:inverse_problem_optim}.
The log-likelihood gradient is finally obtained as:
\begin{equation}\label{eq:likelihood}
    \likelihood \approx - \zeta(\tau) \nabla_{\state} 
\mathcal{C}(\reverb, \rir^{(\psi)}_{\hat{\hrtf}_0}  \ast \speech ),
\end{equation}
where $\zeta(\tau)$ adjusts the weight of the log-likelihood gradient during sampling, 
an is parameterized following \cite{moliner_solving_2022, moliner2024buddy}.
In conclusion, the posterior sampling procedure amounts to solving the following ODE:
\begin{equation} \label{eq:posterior-ode}
\D \state = -
\tau \left[ 
\mathbf{s}_\theta(\state, \tau, \doa)
- \zeta(\tau) \nabla_{\state} 
\mathcal{C}(\reverb, \rir^{(\psi)}_{\hat{\hrtf}_0}  \ast \speech ) 
\right]  \D \tau, 
\end{equation}

In summary, the resulting algorithm alternates between optimizing \ac{brir} parameters and estimating the \ac{hrtf}, as illustrated in  Fig.~\ref{fig:diagram}. At each step $n$ of the discretized diffusion time axis, we perform $N_\mathrm{its.}$ optimization iterations of the parameters $\psi_n$ in our \ac{brir} model \eqref{eq:brir-model}, followed by a sampling step of the ODE \eqref{eq:posterior-ode} to update the \ac{hrtf} estimate $\hrtf_n$.

%% file: sections/exp.tex
\subsection{Experimental Data}
\label{sec:data}

\noindent\textit{HRTF data:}
We obtain the time-aligned features required to train the score model and evaluate our \ac{hrtf} estimation method from the simulated \ac{hrtf} sets of the HUTUBS database~\cite{brinkmann2019a}.
In practice, the pure delay component is estimated and removed from the channel of each \ac{hrtf} data point.
This yields
$2\times 128$-dimensional binaural time-aligned \ac{hrtf} spectra after dropping the Nyquist bin
(more details in~\cite{nam2008method,thuillier2024hrtf}).
Out of the 95 HUTUBS subjects, 85 are used to train the score model.
Two subjects are reserved for validation and six for testing.
Repeated simulations ``88'' and ``96'' of HUTUBS subjects ``1'' and ``22'' are excluded from our splits.

\noindent\textit{Estimation task data:} We evaluate the performance of our HRTF estimation method using reverberant binaural speech observations generated by filtering utterances from VCTK's speakers ``p226'' and ``p287''~\cite{valentini2016reverberant} with simulated \acp{brir}.
The \acp{brir} are generated using a publicly available shoebox simulation software implementing the image-source method~\cite{barumerli2021sofamyroom, schimmel2009shoebox},  which we set to a reflection order of 20.
We provide the software with our test \acp{hrtf} for binauralization of the room impulse response.
Approximately 100 tasks were generated per test subject for a total of 599.

In each estimation task, the room's height is drawn uniformly in the $[2.5,4]$ m range and the floor dimensions (width and length) in the $[7, 15]$ m range.
The source and the subject's head locations is drawn uniformly within the volume of the room at a distance of at least 1.5 m from the walls and with a height in the $[1,2]$ m range.
The source is maintained at least 1 m away from the subject's head and its location is slightly adjusted so that it matches the closest \ac{doa} from the subject's HRTF set.
Finally, the absorption coefficient of the simulation model is drawn within the $[0.05, 0.1]$ range.
A similar validation set was generated for tuning the hyper-parameters of our method using the HUTUBS subjects from our validation split.
The VCTK recordings were down-sampled to 44.1\,kHz prior to filtering so as to match the simulation's sample rate.

 \input{tables/baselines_44k.tex}

\subsection{Implementation Details}

\noindent\textit{Diffusion Parameterization} 
\label{sec:hyperparameters-diffusion}
The score model is trained using diffusion times between
$T_\mathrm{min}=0.01$ and  $T_\mathrm{max}=10$.
At inference time, we reduce the extremal times to $T_\mathrm{min}=0.05$ and  $T_\mathrm{max}=8$ as they showed to be sufficient.
The above levels are relative to normalized HRTF features, which are obtained by scaling the original features with respect to the mean and variance of our training set.
We discretize the diffusion time axis into $N=100$ steps for reverse diffusion using the logarithmic discretization in \cite{karras2022elucidating}.

\vspace{0.5em}
\noindent\textit{DNN Architecture}:
We parameterize the HRTF score model with a 1D-UNet
comprising seven encoding-decoding stages, each with a resampling factor of 2 and comprising 32 hidden features.
The model encodes the \ac{doa} $\gamma$ %
and the diffusion time $\tau$ using random Fourier features embedding~\cite{tancik2020fourier}.
The above architecture results in a total of 752k parameters which we
optimize using Adam with a learning rate of $5\times 10^{-4}$ and batch size of 32 for 110k steps.
During training, we prevent the model from relying too heavily on \ac{doa} information by injecting white noise ($\sigma=0.05$) in the value of $\gamma$, and even completely dropping out $\doa$ with a probability of 30\%.
We track an exponential moving average of the DNN weights with a decay of 0.999.

\vspace{0.5em}
\noindent\textit{BRIR Model}:
STFTs are computed using a Hann window of $23$ ms and $75 \%$ overlap.
We set the number of STFT frames of our reverberation operator to $M_\mathbf{r} = 200$, which corresponds to $120$ ms.
We decimate the frequency scale into $B=40$ bands using a quasi-logarithmic spacing \cite{lemercier2024buddyjournal}.
We optimize the \ac{brir} parameters $\psi$ with Adam \cite{kingma2015adam}, with a learning rate of 0.01, momentum parameters of $\beta_1=0.9$ and $\beta_2=0.999$, and $N_\text{its.}=50$ optimization iterations per diffusion step.
After each optimization step, we further clamp the weights $\left[w_b\right]_b$ between 0 and 40\;dB, and the decays $\left[\alpha_b\right]_b$ between 0.01 and 40. This helps stabilize the optimization at early sampling stages.
The parameters are initialized to $g=0.15$, $t_\mathrm{left}=52$ samples, $w_b=2$ and $\alpha_b=0.1$ (across all frequency bands). Finally, $t_\mathrm{itd}$ is initialized according to the \ac{doa}.

\vspace{0.5em}
\subsection{Evaluation}
We report the performance of our method in terms of logarithmic relative error (LRE)
\begin{align}
	\text{LRE}\left(\hrtf_{c, f}, \hat{\hrtf}_{c, f}\right) &= 20\log_{10} \left|\frac{\hat{\hrtf}_{c, f} - \hrtf_{c, f}}{\hrtf_{c, f}}\right|,\label{eq:relative_error}
\end{align}
and log-magnitude distance (LMD)
\begin{align}
	\text{LMD}\left(\hrtf_{c, f}, \hat{\hrtf}_{c, f}\right) &= \left|20 \log_{10} \left|\frac{\hat{\hrtf}_{c, f}}{\hrtf_{c, f}}\right|\right|.\label{eq:log_magnitude_distance}
\end{align}

We define three baselines for HRTF estimation:
\textit{Random} returns an HRTF filter drawn randomly from the training set at the specified \ac{doa}.
\textit{Generic} 
systematically returns 
\ac{hrtf} filters from the HRTF subject forming a centroid of the training set, i.e. which minimizes the pairwise error to all the other HRTF subjects of the set as computed using the mean LRE across \acp{doa}, frequencies and binaural channels.
\textit{Nearest Neighbour} is a quasi-oracle baseline: 
it selects, amongst the training set, the HRTF with matching \ac{doa} that yields the lowest mean LRE error to the true HRTF.

%% file: tables/baselines_44k.tex
 \begin{table}[t]
    \centering
    \caption{\centering\textit{Instrumental results obtained on simulated 44.1-kHz BRIR data. Values indicate mean and standard deviation. Lower is better}}
    \scalebox{1.0}{
    \begin{tabular}{l|cc}
    
\toprule 
Method & LRE & LMD \\

\midrule
\midrule

Random  & -3.43 $\pm$ 1.59 & 4.58 $\pm$ 0.69 \\

Generic & -4.01 $\pm$ 1.75 & 4.77 $\pm$ 0.74 \\

Proposed  & \textbf{-9.20} $\pm$ \textbf{5.64} & \textbf{2.28} $\pm$ \textbf{1.11} \\

\midrule

\textit{Nearest Neighbour} & \textit{-7.50 $\pm$ 1.46} & \textit{3.76 $\pm$ 0.61} \\

 \midrule
    \bottomrule
    \end{tabular}
    }
    \label{tab:baselines-44k}
\end{table}

%% file: figures/plots_v3.tex
\newcommand{\gw}[0]{0.235\textwidth}
\newcommand{\gh}[0]{3.15cm}
\newcommand{\gs}[0]{0.43}
\begin{figure}[t]
    \begin{center}
    \hspace{0.7cm}
    \includegraphics[width=0.30\textwidth]{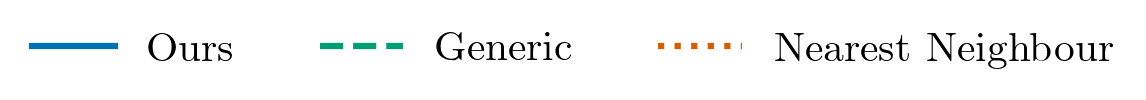}
    \end{center}
    \vspace{-0.2cm}
    \hspace{0.cm}
    \begin{subfigure}[b]{0.245\textwidth}
        \centering
        \includegraphics[height=\gh]{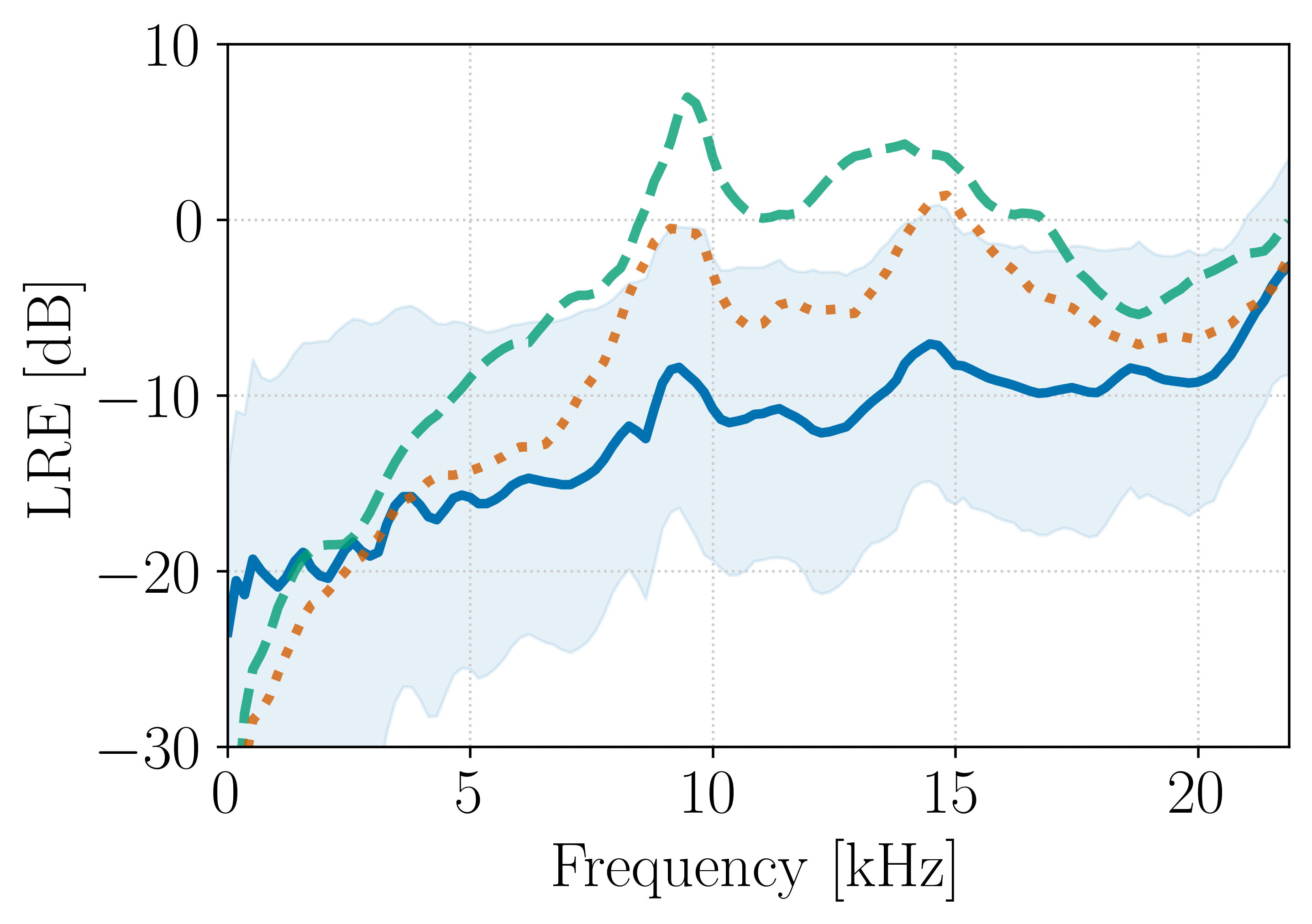}
           \captionsetup{justification=centering, margin=2.5cm}
        \vspace{-0.55cm}
          \caption{} %
    \end{subfigure}%
    \begin{subfigure}[b]{\gw}
        \centering
        \includegraphics[height=\gh]{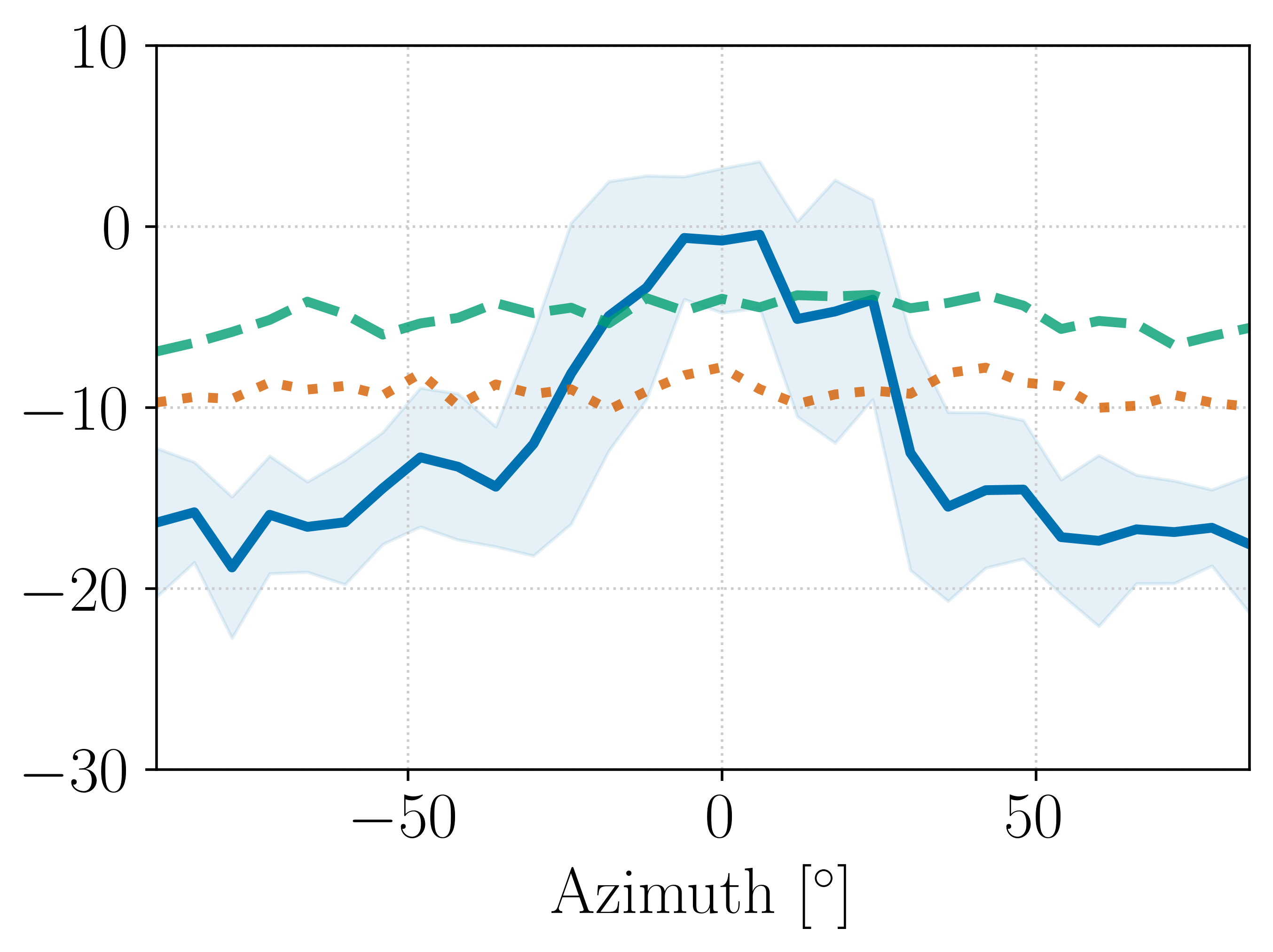}
           \captionsetup{justification=centering, margin=2.23cm}
        \vspace{-0.55cm}
        \caption{}
    \end{subfigure}%
    
    \caption{\textit{LRE as a function of (a) frequency and (b) azimuth.
    }}
    \label{fig:plots}
\end{figure}

\begin{figure}[t]
    \centering
    \begin{center}
    \hspace{0.2cm}
    \includegraphics[width=0.30\textwidth]{images/44k/legend.png}
    \end{center}
    \vspace{-0.2cm}
    \begin{subfigure}[b]{0.245\textwidth}
        \centering
        \includegraphics[height=\gh]{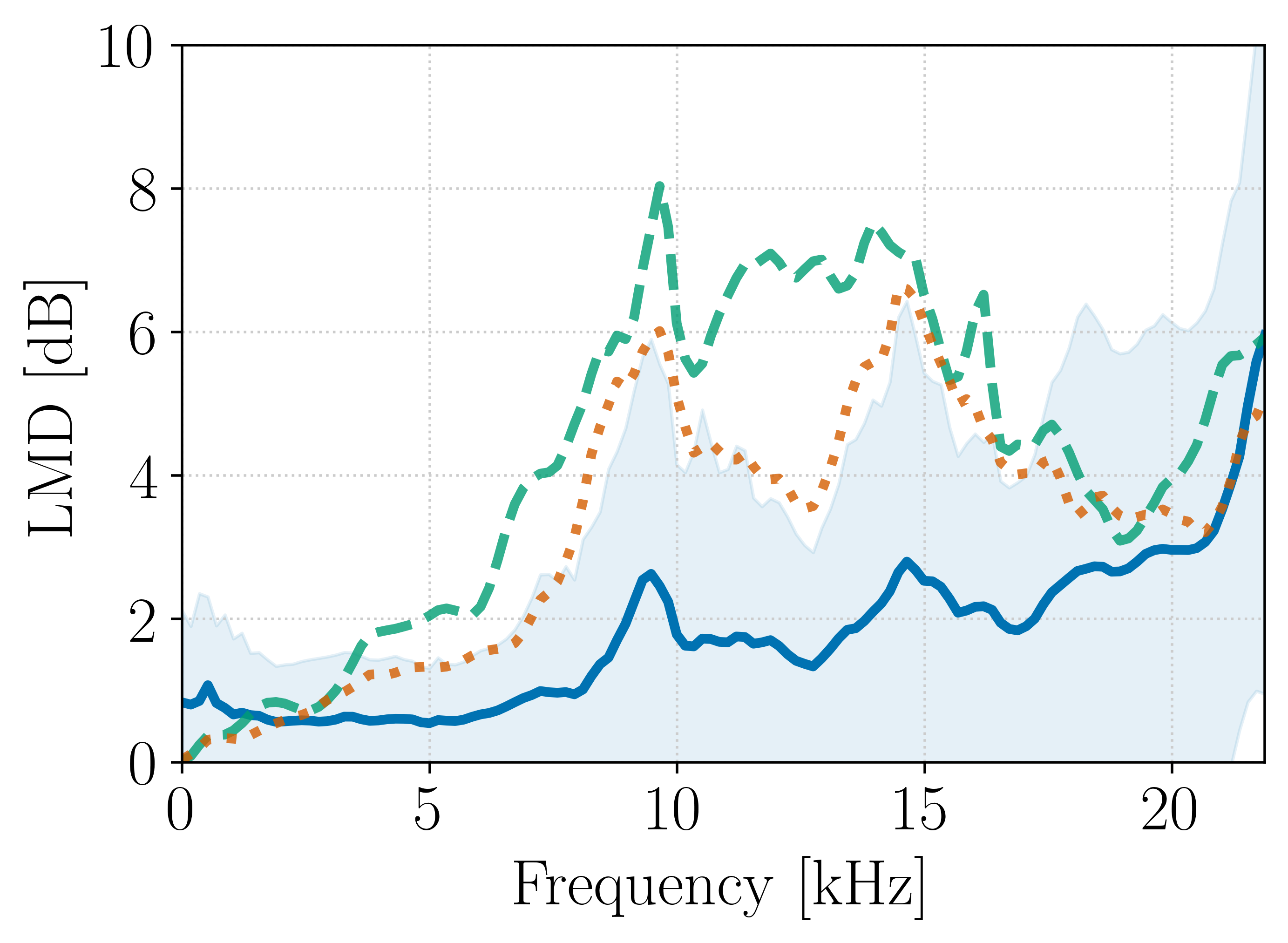}
           \captionsetup{justification=centering, margin=2.35cm}
        \vspace{-0.15cm}
          \caption{} %
    \end{subfigure}%
    \begin{subfigure}[b]{\gw}
        \centering
        \includegraphics[height=\gh]{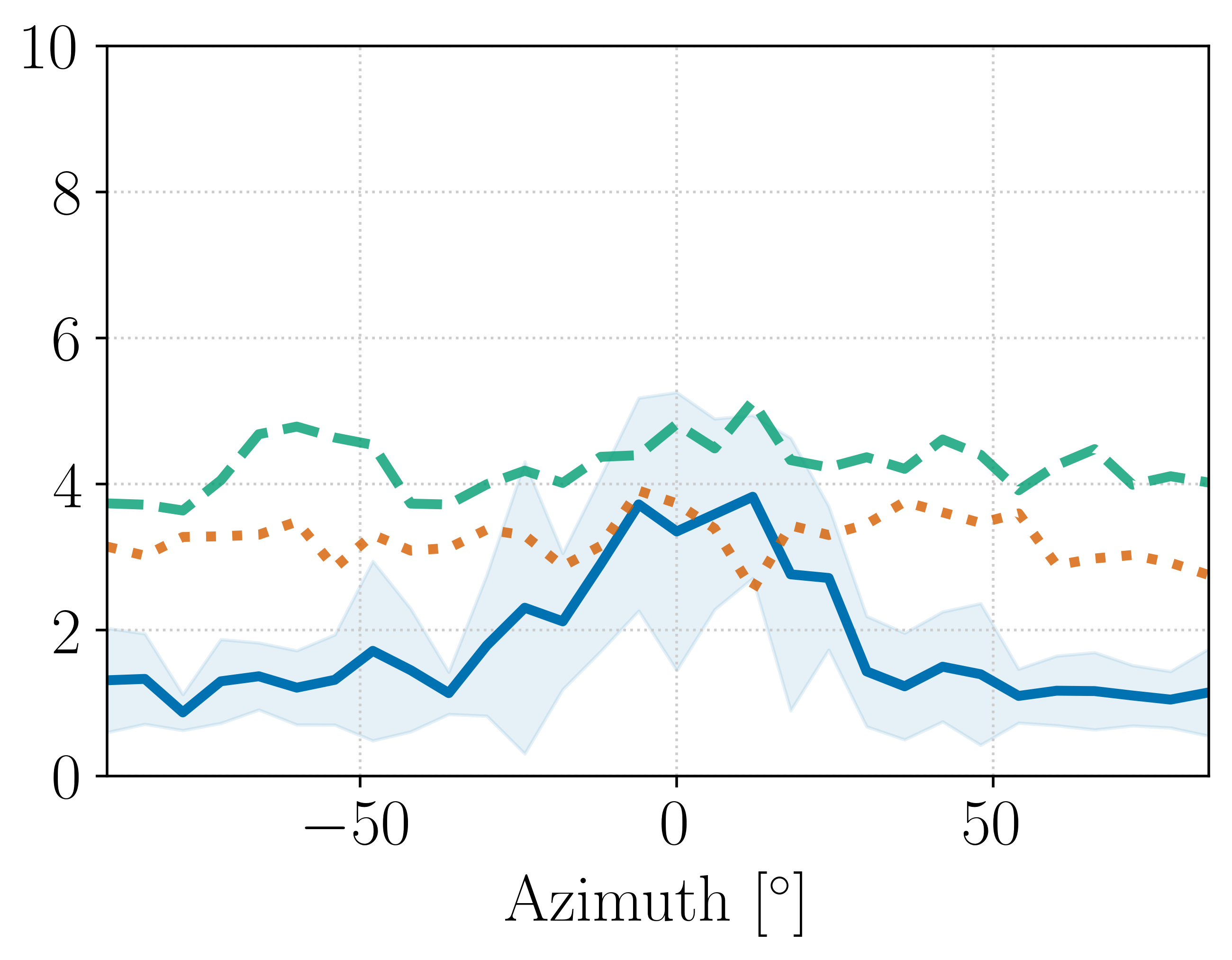}
           \captionsetup{justification=centering, margin=2.1cm}
        \vspace{-0.15cm}
        \caption{}
    \end{subfigure}%
    
    \caption{\textit{LMD as a function of (a) frequency and (b) azimuth.
    }}
    \label{fig:plots-lmd}
\end{figure}

%% file: sections/results.tex
The results of the objective metrics are reported in Table \ref{tab:baselines-44k}. The proposed method outperforms the compared baselines, including the \textit{Nearest Neighbour} oracle in both LRE and LMD metrics. 
Figures \ref{fig:plots} and \ref{fig:plots-lmd} illustrate the instrumental metrics as a function of frequency and azimuth. 
The results in Figures \ref{fig:plots}a and \ref{fig:plots-lmd}a reveal several key trends. First, the error increases with frequency, likely due to higher individual variability at higher bands. 
Notably, in the 5–8 kHz range, our method achieves a mean LRE that is at least 8 dB lower than the \textit{Generic} HRTF baseline.
Furthermore, in the higher frequency range (8–17 kHz), the proposed method improves over the \textit{Nearest Neighbour} oracle baseline by at least 6 dB in LRE and 2 dB in LMD. This performance gain highlights that our method surpasses mere data retrieval capabilities, which we attribute to the modeling capacity of the score-based prior.
At lower frequencies (0–1 kHz), the error is slightly higher than the \textit{Generic} baseline and \textit{Nearest Neighbour}. 
However, this occurs below the range in which lie the most salient monaural cues, in particular filtering from the pinna ($>$3 kHz) ~\cite{jin2004contrasting}.

One area of concern is the increased LRE observed in Fig.\;\ref{fig:plots}b at the median plane, i.e. 0$^\circ$ azimuth.
In terms of LMD, the proposed solution also suffers from lower performance at the median plane, but our method still outperforms the \textit{Generic} \ac{hrtf} and is on par with the \textit{Nearest Neighbour} baseline in this worst azimuth case. 
This suggests that the phase estimation (only assessed in the LRE metric) seems to suffer more than the magnitude estimation in this region. This overall phenomenon may be due to inherent challenges in modeling HRTFs at this spatial position, but this warrants further investigation.